\documentclass[aps,prl,amsmath,amsfonts,amssymb,twocolumn,superscriptaddress,showpacs]{revtex4-1}

\usepackage{graphicx,psfrag,color}
\usepackage{dcolumn}
\usepackage{bm}
\usepackage{mathbbol,dsfont}

\begin{document}

\title{The Adiabatic Theorem for Quantum Systems with Spectral Degeneracy}

\author{Gustavo Rigolin}
\affiliation{Departamento de Fisica,
Universidade Federal de Sao Carlos, Sao Carlos, SP 13565-905,
Brazil}
\author{Gerardo Ortiz}
\affiliation{Department of Physics, Indiana University,
Bloomington, IN 47405, USA}
\date{\today}

\begin{abstract}
By stating the adiabatic theorem of quantum mechanics in a clear
and rigorous way, we establish a necessary condition and 
a sufficient condition for its validity, where the latter is obtained
employing our recently developed adiabatic perturbation theory. Also,
we simplify further the sufficient condition into a useful and simple
practical test at the expenses of its mathematical rigor. We 
present results for the most general case of quantum systems,
i.e., those with degenerate energy spectra. These conditions are of 
upmost importance to assess the validity of practical implementations 
of non-Abelian braiding and adiabatic quantum computation. 
To illustrate the degenerate adiabatic
approximation, and the necessary and sufficient conditions for its validity, we 
analyze in depth  an exactly solvable time-dependent degenerate problem. 
\end{abstract}

\maketitle

The adiabatic theorem \cite{Mes62} has played, and still plays, a fundamental role 
in practical quantum physics applications. Indeed, the ability to determine
how the {\it slow} dynamics of external probes coupled to a system affect 
its time evolution has applications
ranging from the notion of thermal equilibrium and non-equilibrium phenomena \cite{Pol11}
to the conditions under which
an adiabatic quantum computer can reliably operate \cite{Joh11}. Useful and practical 
quantitative conditions for the validity of the 
adiabatic theorem are also relevant to the important current problem of assessing 
the feasibility of any information processing scheme that uses the concept of fractional
exchange statistics  and non-Abelian braiding \cite{Nay08}. 

General physical principles dictate that, in three space dimensions, 
elementary particles can only obey  fermionic or bosonic statistics. 
Kinematic constraints do not allow for fractional exchange statistics: electrons are spin-1/2 fermions 
and photons are spin-1 bosons. 
Nonetheless, fractional statistics {\it particles} or {\it modes} may emerge from the collective 
behavior of elementary particles, i.e., collective excitations of a quantum field, 
as a result of a dynamical process. 
The latter requires special circumstances and constraints 
that should be analyzed on a case by case basis. 
For instance, for two localized degenerate Majorana modes to realize 
a non-Abelian braiding process we need to design the
physical Hamiltonians realizing the braiding that do not lift the 
degeneracy and can be implemented adiabatically.
If those constraints are not met
experimentally then  the braiding operation is faulty. 
Physical systems where such fractional statistics emerges  have
highly degenerate energy spectrum, thus justifying a careful 
statement of the adiabatic theorem and the precise conditions for its validity.

Despite its practical importance, 
no consensual and rigorous necessary and sufficient conditions for the validity
of the adiabatic theorem have been given.
Only recently a proof that the 
commonly used textbook condition \cite{Mes62} is necessary 
for non-degenerate Hamiltonians \cite{Ton10} but not sufficient \cite{Ton05} was given. 
For degenerate systems,
even a clear presentation of the theorem is lacking, let alone 
necessary and sufficient conditions. It is this paper's intention to 
fill that gap. 

With that in mind, our goal is three fold. 
First, using techniques developed in \cite{Rig08,Rig10}, 
we aim at providing a clear and rigorous version of the adiabatic theorem 
for Hamiltonians with non-degenerate and degenerate spectra using a single formalism. 
We want to be as precise as possible 
in stating the adiabatic theorem  to avoid common misunderstandings \cite{Ton10a},
mainly due to a lack of quantitative rigor in the 
way the theorem is usually presented. Second, we prove necessary and sufficient
conditions for the validity of the rigorous version of the adiabatic theorem
here presented. The necessary condition for degenerate spectra reduces to the one in 
\cite{Ton10} when no degeneracy is present. To obtain a sufficient condition, we rely
on the adiabatic perturbation theory developed in \cite{Rig08,Rig10}. Finally, we apply
these ideas to an exactly solvable time-dependent degenerate problem \cite{Rig10}, where
we show that the necessary and sufficient conditions here developed provide the
correct conditions under which the adiabatic theorem holds.   

To properly formulate the degenerate adiabatic theorem (DAT) we first need to 
introduce the degenerate adiabatic approximation (DAA). As we will see,
DAT is essentially a  statement about the mathematical conditions for the validity of
DAA. This understanding of the essence of the adiabatic theorem is akin to
the ones of Berry \cite{Ber84} and Tong \cite{Ton10}, for non-degenerate systems,
and to the ones of Wilczek and Zee \cite{Wil84} and Wilczek \cite{Wil11},
for degenerate systems. 

{\it Degenerate Adiabatic Approximation.} Consider an explicitly time-dependent Hamiltonian
$\mathbf{H}(t)$ with orthonormal eigenvectors $|n^{g_n}(t)\rangle$, where
$g_n=0,1,\cdots,d_n-1$ labels states of the degenerate eigenspace $\mathcal{H}_n$ of 
dimension $d_n$ and eigenenergy $E_n(t)$, 
$\mathbf{H}(t)|n^{g_n}(t)\rangle = E_n(t)|n^{g_n}(t)\rangle$;
and assume that $d_n$ does not change during the total time 
evolution, $t\in[0,T]$. An arbitrary state at $t=0$
can be written as 
$|\Psi^{(0)}(0)\rangle=\sum_n\sum_{g_n=0}^{d_{n-1}}b_n(0)U_{h_ng_n}^n(0)|n^{g_n}(0)\rangle$,
where $|b_n(0)|^2$ gives the probability of the system being in eigenspace 
$\mathcal{H}_n$ and $|b_n(0)U_{h_ng_n}^n(0)|^2$ the probability of measuring 
a specific eigenstate. 
A given initial condition
within an eigenspace is characterized by one value of $h_n=0,1,\cdots,d_n-1$. 
A compact way of representing all possible initial conditions spanning the orthonormal 
eigenspace $\mathcal{H}_n$ is  \cite{Rig10},
$\mathbf{|\Psi}^{(0)}(0)\rangle = \sum_{n=0}
b_n(0)\mathbf{U}^{n}(0)\mathbf{|n}(0)\rangle$, where 
$\mathbf{|n}(t)\rangle$ $=$ $(|n^0(t)\rangle,$ $|n^1(t)\rangle,$ $\ldots,$ $|n^{d_n-1}(t)\rangle)$
is a column vector, and $\mathbf{U}^{n}(0)$ a $d_n\times d_n$ unitary matrix, 
$\mathbf{U}^{n}(0)(\mathbf{U}^{n}(0))^\dagger = \mathds{1}$.
A particular initial state corresponds to choosing the  
corresponding element of the column vector $\mathbf{|\Psi}^{(0)}(0)\rangle$.

Then, the most general way of writing DAA is
\begin{equation}
\mathbf{|\Psi}^{(0)}(t)\rangle = \sum_{n=0}
\mathrm{e}^{-\mathrm{i}\omega_n(t)}
b_n(0)\mathbf{U}^{n}(t)\mathbf{|n}(t)\rangle, \label{vector0}
\end{equation}
where 
%
$\omega_n(t) \!=\! \int_{0}^{t}E_n(t')dt'/\hbar$
%
is the dynamical phase, and the 
unitary matrix 
$\mathbf{U}^{n}(t) = \mathbf{U}^{n}(0)\mathcal{T}
\exp( \int_0^t\mathbf{A}^{nn}(t')dt')$ the non-Abelian
Wilczek-Zee phase (WZ phase).
%
%
Here $\mathcal{T}$ denotes a time-ordered operator, and
$A^{nm}_{h_ng_m}(t)=(M^{nm}_{h_ng_m}(t))^*$ a 
$d_n \times d_n$ matrix defined as
\begin{equation}
[\mathbf{M}^{mn}(t)]_{g_mh_n}=M^{nm}_{h_ng_m}(t)=\langle n^{h_n}(t)| \dot{m}^{g_m}(t)\rangle,
\label{M}
\end{equation}
with the dot meaning time derivative. For example, for a
system starting at the ground eigenspace ($b_n(0)= \delta_{n0}$)
$\mathbf{|\Psi}^{(0)}(t)\rangle=
\mathrm{e}^{-\mathrm{i}\omega_0(t)} 
\mathbf{U}^{0}(t)\mathbf{|0}(t)\rangle$.


The time evolution of an informationally isolated quantum system
 is dictated by the Schr\"odinger equation (SE)
$\mathrm{i}\,\hbar\, |\dot{\mathbf{\Psi}}(t)\rangle =
\mathbf{H}(t) |\mathbf{\Psi}(t)\rangle.
$
What are the constraints on the rate of change of $\mathbf{H}(t)$
under which the system's evolved state $|\mathbf{\Psi}(t)\rangle$ 
gets {\it close} to DAA? The adiabatic theorem we formulate next 
sets  the conditions under which DAA
holds. In other words, it precisely states when the system's dynamics 
can be approximated by DAA.

{\it Adiabatic Theorem:} If a system's Hamiltonian $\mathbf{H}(t)$ changes \textit{slowly} 
during the course of
time, say from $t=0$ to $t=T$, and the system is prepared in an arbitrary superposition
of eigenstates of $\mathbf{H}(t)$ at $t=0$, say
$|\mathbf{\Psi}^{(0)}(0)\rangle$,
then the transitions between eigenspaces
$\mathcal{H}_n$ of $\mathbf{H}(t)$ during the interval $t\in [0,T]$ are \textit{negligible} 
and the system \textit{evolves} according to DAA.

The three important concepts, {\it slow, negligible},  and {\it evolved state}, 
need further explanation. 
First, DAA is based on the assumption that the 
rate of change of $\mathbf{H}(t)$ is {slow}. A crucial matter is then to 
establish the meaning of slow precisely.  Intuitively, the 
latter notion can be understood as a relation between a characteristic {\it internal} 
time of the evolved system $T_i$, encoded in $\mathbf{H}(t)$,  
and the total evolution time $T$, such that $T_i/T\ll 1$. For a fixed 
and finite $T_i$, one can always choose an evolution time $T$ that 
satisfies this condition. This state of affairs, however, is not satisfactory 
from a mathematical standpoint. Indeed, a main source of controversy in the 
literature arises from the lack of a precise  
quantification of the term {\it slow}. By using the degenerate 
adiabatic perturbation theory (DAPT) \cite{Rig10}, a generalization of
APT \cite{Rig08}, we can give a precise meaning to this notion of 
slowness, which is the key ingredient to the derivation of the sufficient condition
of DAT. 
Second, to establish the necessary condition we follow Tong \cite{Ton10} and others 
\cite{Ber84,Wil84,Wil11} and assume that if the system's state is well described by DAA
then all measurements performed on the system at {\it any} time must indeed be consistent with this assumption. 
This has a profound implication on the approximate dynamics the system obeys \cite{Ton10}.
The following necessary and sufficient conditions provide the mathematical 
rigor required to make those concepts precise.

{\it The necessary condition.} There is no unique way of establishing 
how  {\it close}  two quantum 
states are, implying that there is no unique distance measure 
between states. A popular choice in the 
context of quantum information is the fidelity measure. We  
stress though that DAT 
is not a statement about the fidelity between the true 
time-dependent state $|\Psi(t)\rangle$ and DAA 
$|\Psi^{(0)}(t)\rangle$ being close to one, i.e., 
$|\langle \Psi(t) | \Psi^{(0)}(t) \rangle | \sim 1$.
It is more than that, it is a statement about  DAA 
expectation value of {\it any} observable 
being {\it close} to the exact ones. 
This notion is crucial to define geometric phases, 
thus for particle exchange statistics, and is crucial for the philosophy behind DAPT
and the proof of necessity that now follows.  

If DAA is an accurate description of
the time evolution of a degenerate system starting, with no loss of generality,
in its ground eigenspace ($b_n(0)=\delta_{n0}$)
then $\mathbf{|\Psi}(t)\rangle$ $=$ $\mathbf{|\Psi}^{(0)}(t)\rangle$ $+$ ${\cal O}(1/T) 
\approx \mathbf{|\Psi}^{(0)}(t)\rangle$, 
with $||{\cal O}(1/T)||_{\text{max}}\ll 1$, where $|| \cdot ||_{\text{max}}$ is the max
norm (the absolute value of the greatest element of a given vector/matrix). 
It immediately
follows that the system (a) approximately satisfies SE
$\mathrm{i}\,\hbar\, |\mathbf{\dot{\Psi}}^{0}(t)\rangle \approx
\mathbf{H}(t) |\mathbf{\Psi}^{0}(t)\rangle 
$
which \textit{implies} \cite{Ton10} 
$|\mathbf{\dot{\Psi}}(t)\rangle \approx |\mathbf{\dot{\Psi}}^{(0)}(t)\rangle;$
and that (b) transitions to excited \textit{eigenspaces} are negligible \cite{footnote1},
$\left\|\mathbf{\langle n}(t)|^T\mathbf{|\Psi}(t)\rangle^T\right\|_\text{max} \ll 1,
n \neq 0.
$

Now, using (a), (b), and defining $\Delta_{nm}(t)=E_n(t)-E_m(t)$ 
we notice that for $n \neq 0$ \cite{Rig12},
%
\begin{eqnarray*}
\mathbf{\langle n}(t)|^T\mathbf{|\Psi}(t)\rangle^T&=& 
\frac{\mathbf{\langle n}(t)|^T(\mathbf{H}(t) - E_0(t))\mathbf{|\Psi}(t)\rangle^T}{\Delta_{n0}(t)} \nonumber \\
&=& \frac{\mathbf{\langle n}(t)|^T\left(\mathrm{i}\hbar\mathbf{|\dot{\Psi}}(t)\rangle^T
- E_0(t)\mathbf{|\Psi}(t)\rangle^T\right)}{\Delta_{n0}(t)} \nonumber \\
&\approx& \frac{\mathrm{i}\hbar\mathbf{\langle n}(t)|^T\mathbf{|\dot{\Psi}}^{(0)}(t)\rangle^T}{\Delta_{n0}(t)} 
\nonumber \\
&=& \mathrm{i}\hbar\mathrm{e}^{-\mathrm{i}\omega_0(t)}
\frac{\mathbf{\langle n}(t)|^T[\mathbf{U}^{0}(t)\mathbf{|\dot{0}}(t)\rangle]^T}{\Delta_{n0}(t)},
\end{eqnarray*}
where $\mathbf{\langle n}(t)|^T\mathbf{|0}(t)\rangle^T=\mathbf{0}$.  
%
Taking the max norm on both sides and using (b)
we get the necessary condition 
%
$\hbar\left\|\mathbf{\langle n}(t)|^T[\mathbf{U}^0(t)\mathbf{|\dot{0}}(t)\rangle]^T
/\Delta_{n0}(t)\right\|_{\sf max} \hspace{-.25cm}\ll 1, \hspace{.2cm} n \neq 0, \hspace{.2cm} t \in [0,T]$.
%
Finally, using that $\|\mathbf{U}^{n}(t)\|_{\sf max}\leq1$ leads
to a stronger WZ phase-free  necessary condition,
\begin{equation}
\hbar\left\|\frac{\mathbf{M}^{n0}(t)}{\Delta_{n0}(t)}\right\|_1
\ll 1, \hspace{.5cm} n \neq 0, \hspace{.5cm} t \in [0,T],
\label{strongerMat}
\end{equation}
where $\left \| A \right \|_1 = \max \limits_{1 \leq j
\leq {p}} \sum _{i=1}^{q} | a_{ij} |$ for a $p\times q$
dimensional matrix $A$. When the spectrum is non-degenerate 
($d_n=1$), Eq. (\ref{strongerMat}) reduces to the 
necessary condition of Ref. \cite{Ton10}.

{\it The sufficient condition.}
The first stept to stablish the sufficient condition 
is to prove the convergence of DAPT
in its full generality. Intrinsic to the formulation of
DAPT is a Taylor series expansion in terms of the parameter $v=1/T$, 
and a necessary rescaling of time according to $s=vt$ with 
$s\in [0,1]$  \cite{Rig10}. For small enough $v$ one can always make
DAPT converge (cf. Eq. (\ref{strongerSuf})). 

Inserting the ansatz 
\begin{eqnarray}
|\mathbf{\Psi}(s)\rangle 
&=& \sum_{n=0}\sum_{p=0}^{\infty}
\mathbf{C}_{n}^{(p)}(s)|\mathbf{n}(s)\rangle
\end{eqnarray}
into SE 
with
$\mathbf{C}_{n}^{(p)}(s) =
\mathrm{e}^{-\frac{\mathrm{i}}{v}\omega_n(s)}v^p
\mathbf{B}_{n}^{(p)}(s)$ and 
$
\mathbf{B}_{n}^{(p)}(s)=
\sum_{m=0}
\mathrm{e}^{\frac{\mathrm{i}}{v}\omega_{nm}(s)}
\mathbf{B}_{mn}^{(p)}(s),
$
DAPT gives recursive equations for $\mathbf{B}_{mn}^{(p)}(s)$
in terms of lower order in $p$ coefficients \cite{Rig10}. The zeroth order
is exactly DAA, with WZ phase naturally appearing as a requirement
for the consistency of the series expansion.
Note that for each $n$ we have a series involving the matrix
$\mathbf{C}_{n}^{(p)}(s)$, $p=0,1,\ldots, \infty$. The matrix
element $[\mathbf{C}_{n}^{(p)}(s)]_{h_ng_n}$ is the coefficient
giving the contribution to order $p$ of the state
$|n^{g_n}(s)\rangle$ to the solution to SE. Here $h_n$ handles different initial conditions and
for definiteness we pick the case $h_n=0$, $\forall n$. 
Applying the ratio test for series expansions, if the 
condition
\begin{equation}
\lim_{p\rightarrow
\infty}\left|[\mathbf{C}_{n}^{(p+1)}(s)]_{0g_n}
/[\mathbf{C}_{n}^{(p)}(s)]_{0g_n}\right| < 1, \hspace{.5cm}
\forall n, g_n, \label{cn}
\end{equation}
is satisfied for all coefficients then 
we guarantee convergence of DAPT.
We can simplify further (\ref{cn}) by invoking the
comparison test \cite{Rig12},
%
\begin{equation}
\lim_{p\rightarrow \infty}\frac{v\sum_{m=0}
\left|\left[\mathbf{B}_{mn}^{(p+1)}(s)\right]_{0g_n}\right|}
{\sum_{m=0} 
\left|\left[\mathbf{B}_{mn}^{(p)}(s)\right]_{0g_n}\right|} < 1,
\hspace{.5cm} \forall n, g_n.
\label{strongerSuf}
\end{equation}

Imposing that $\sum_{p=0}^{\infty}|[\mathbf{C}_{n}^{(p+1)}(s)]_{0g_n}|
\ll |[\mathbf{C}_{n}^{(0)}(s)]_{0g_n}|$, $\forall n, g_n $, 
meaning that the zeroth order dominates, is equivalent to  
\begin{equation}
\sum_{p=0}^{\infty}\sum_{m=0}v^{p+1} 
\left|\left[\mathbf{B}_{mn}^{(p+1)}(s)\right]_{0g_n}\right|\ll
\sum_{m=0}
\left|\left[\mathbf{B}_{mn}^{(0)}(s)\right]_{0g_n}\right|, 
\label{suf}
\end{equation}
which together with Eq. (\ref{strongerSuf}) are the
rigorous sufficient conditions for the validity of
DAA. In practice  it is
extremely difficult to compute the previous limit when $p
\rightarrow \infty$ and all orders $p$. We can come up, nevertheless, 
with some practical condition of convergence 
by looking at the ratio for a couple of finite orders
$p$.  Working with increasing $p$ we get more and more
conditions that, in the non-degenerate case, can become stronger than
the ones in \cite{Ton07}. In its simplest form, we may consider only $p=0$.
 In this case both expressions merge into one and we demand it to
be \textit{much smaller} than 
the smallest \textit{non-null} term appearing in the rhs of (\ref{suf}). 
Thus, the practical sufficient test reads 
\begin{equation}
v\sum_{m=0}
\left|\left[\mathbf{B}_{mn}^{(1)}(s)\right]_{0g_n}\right|\ll
\min \limits_{n,g_n}\sum_{m=0}
\left|\left[\mathbf{B}_{mn}^{(0)}(s)\right]_{0g_n}\right|.
\end{equation}

Using  \cite{Rig10}
$\mathbf{B}_{mn}^{(0)}(s) = b_n(0)\mathbf{U}^n(s)\delta_{mn}$
and the fact that at $t=0$ the initial state is $|0^0(0)\rangle$ 
($b_n(0)=\delta_{n0}$) we get    
\begin{equation}
v\hspace{-.1cm}\sum_{m=0}\hspace{-.05cm}
\left|\hspace{-.05cm}\left[\mathbf{B}_{mn}^{(1)}(s)\right]_{0g_n}\hspace{-.05cm}\right|
 \ll
\min \limits_{g_0}
\left(\left|\left[\mathbf{U}^{0}(s)\right]_{0g_0}\right|\right),
\forall n,g_n, 
\label{suf3}
\end{equation}
which is
our intuitive and practical sufficient condition. Indeed, noting that
$v$ $\sum_{m=0}$ $\mathrm{e}^{-\frac{\mathrm{i}}{v}\omega_m(s)}$ 
$\left[\mathbf{B}_{mn}^{(1)}(s)\right]_{0g_n}$,
with $n\neq 0$, gives the first order contribution of the excited
state $|n^{g_n}(s)\rangle$ to the wave equation, and that for $n=0$ it 
is related to the first order correction to the WZ phase \cite{Rig10}, 
it is clear that they must be much smaller than the smallest coefficient
appearing in the zeroth order if we want DAA to hold. 

Equation (\ref{suf3}) also depends on
$\mathbf{U}^{n}(s)$  because
$\mathbf{B}_{mn}^{(1)}(s)$ depends on $\mathbf{U}^{n}(s)$.
However, a similar calculation to the one done for 
 the necessary condition gets rid of these
unitary matrices  leading to \cite{Rig12}
\begin{equation}
D_{g_n}^{n}(t) \ll \min \limits_{g_0}
\left(\left|\left[\mathbf{U}^{0}(t)\right]_{0g_0}\right|\right), 
\hspace{.5cm} t\in [0,T], \label{d} 
\end{equation}
where for $n=0$ and $\forall g_0$ we have $D_{g_0}^{0}(t)$ equals to
\begin{equation}
\hbar d_0 \!\!\! \int_0^t\!\!\!\! \mathrm{d}t'\!\sum_{n=1}
\!\!\left\{\frac{\sum_{k_0,i_0=0}^{d_0-1}|[\mathbf{M}^{0n}(t')
(\mathbf{M}^{0n}(t'))^\dagger]_{k_0i_0}|}{|\Delta_{0n}(t')|}\right\},
\label{d1}
\end{equation}
and for $n\neq 0$ and $\forall g_n$, $D_{g_n}^{n}(t)$ is given by
\begin{equation}
\!\frac{\hbar}{|\Delta_{n0}(0)|}\!\!\left\{\! \sum_{k_0=0}^{d_0-1}
\hspace{-.15cm}\left|[\mathbf{M}^{0n}(t)]_{k_0g_n}\!\right| \!+\!
d_n\!\hspace{-.5cm}\sum_{k_0,l_n=0}^{d_0-1,d_n-1}\hspace{-.55cm}
\left| [\mathbf{M}^{0n}(0)]_{k_0l_n}\! \right|\!\!\right\}\!.
\label{d2}
\end{equation}

{\it Example.} We now apply the previous ideas to a doubly degenerate four-level system
subjected to a rotating magnetic field  of constant magnitude
$\mathbf{B}(t) = B \mathbf{r}(t)$ and in spherical
coordinates $ \mathbf{r}(t) = (\sin \theta \cos w\,t, \sin
\theta \sin w\,t, \cos \theta ), $ with $w>0$ and $0 \leq \theta \leq
\pi$ being the polar angle. The Hamiltonian describing this system is
\cite{Bis89,Rig10}
%
$
\mathbf{H}(t) = \hbar b\, \mathbf{r}(t) \cdot \mathbf{\Gamma}/2,
$
%
where $b>0$ is proportional to the coupling between the field and the system
and $\bm{\Gamma}=(\Gamma_x,\Gamma_y,\Gamma_z)$ are the Dirac matrices
$\Gamma_j = \sigma_x \otimes \sigma_j$, $j=x,y,z$. Here $\sigma_j$ are the
standard Pauli matrices implying the following algebra for $\Gamma_j$,
$
\{\Gamma_i,\Gamma_j\} = 2\delta_{ij}\bm{I}_4,
$
$
[\Gamma_i,\Gamma_j] = 2\mathrm{i}\epsilon_{ijk}\Pi_k,
$
where $\bm{I}_4$ is the identity matrix of dimension four,
$\delta_{ij}$ the Kronecker delta,
$\epsilon_{ijk}$ the Levi-Civita symbol, and
$\Pi_k = \bm{I}_2\otimes \sigma_k$.
Starting at the ground state $|0^{0}(0)\rangle$ the 
time-dependent 
solution in terms of the snapshot eigenstates is \cite{Rig10} 
%
$|\Psi(t)\rangle$ $=$ $e^{\mathrm{i} w t/2}$ 
$[ (1 + \cos\theta)A_-(t)$ $+$ $(1-\cos\theta)A_+(t)]/2$ $|0^0(t)\rangle$ 
$+$ $e^{-\mathrm{i} w t/2}\sin\theta$ $( A_+(t) - A_-(t))/2$ 
$|0^1(t)\rangle$ $+$ $e^{\mathrm{i} w t/2}\sin^2\theta$
$( B_+(t) + B_-(t))/2$ $|1^0(t)\rangle$ $+$ $e^{-\mathrm{i} w t/2}$
$\sin\theta$ $[ (1 + \cos\theta)B_-(t)$ $-$ $(1 - \cos\theta) B_+(t)]/2$
$|1^1(t)\rangle$,
%
where
%
$A_\pm(t)$ $=$ $\cos(\Omega_\pm t/2)$ $+$ $\mathrm{i}$ $(b\pm
w\cos\theta)$ $\sin(\Omega_\pm t/2)/\Omega_\pm$, 
$B_\pm(t)$ $=$ $\mathrm{i}$ $w$ $\sin(\Omega_\pm t /2)/\Omega_\pm$,
$\Omega^2_\pm$ $=$ $w^2$ $+$ $b^2$ $\pm$ $2wb\cos\theta$. 

\textit{Necessary condition.} Since in this example
Eq. (\ref{M}) is $[\mathbf{M}^{10}(t)]_{11}=-[\mathbf{M}^{10}(t)]_{00}
=\mathrm{i}w\sin^2(\theta)/2$ and $[\mathbf{M}^{10}(t)]_{10}=-[\mathbf{M}^{10}(t)]^{*}_{01}
=-\mathrm{i}w\sin(2\theta)e^{\mathrm{i}wt}/4$ 
%
%
the necessary condition (\ref{strongerMat}) becomes
%
$w\sin\theta|\sin\theta + \cos\theta|/(2b) \ll 1$.
%
Our task now is to look at the exact solution, impose that DAA holds, 
and see if it implies the necessary condition above. If DAA
holds then the absolute values of the
coefficients multiplying $|1^0(t)\rangle$ and $|1^1(t)\rangle$ must
be negligible. This leads to \cite{Rig12}
$w\sin\theta f(\theta)/(2b) \ll 1$,
with $f(\theta)=|b/\Omega_+ + b/\Omega_- + \cos\theta
(b/\Omega_- - b/\Omega_+)|$. Noting that $f(\theta)$
has a global minimum at $\theta=\pi/2$ equal to $2b/\sqrt{b^2+w^2}$
it is not difficult to see that if $w<b$ than $f(\theta)\geq \sqrt{2}$.
Hence, $1\gg w\sin\theta f(\theta)/(2b)\geq w\sin\theta \sqrt{2}/(2b) \geq 
w\sin\theta |\sin\theta +\cos\theta|/(2b)$, which is exactly
the necessary condition. When $w\geq b$ we have $f(\theta)\geq b\sqrt{2}/w$
which leads to $w\sin\theta f(\theta)/(2b)\geq\sqrt{2}\sin(\theta)/2 \approx \sin\theta$.
Since $\sin\theta \approx 1$  DAA is not a faithful approximation to the exact state
for general $\theta$ when $w\geq b$. This is expected since the rotating frequency
$w$ of the magnetic field must be much smaller than the coupling constant 
$b$ (natural frequency of the system) for DAA to hold. The pathological situation where $\sin\theta \rightarrow 0$
and the fidelity of the state approaches unity 
even though $w\geq b$ does not lead
to a state evolving according to DAA \cite{Ton10a}.

\textit{Sufficient condition.} Equations (\ref{d}-\ref{d2}) become
for $t\in[0,T]$,
%
%
$w^2t\sin^2(\theta)/b \ll \min \limits_{ g_0}
(|[\mathbf{U}^{0}(t)]_{0g_0}|)$
%
and
$5w\sin\theta(|\cos\theta|+\sin\theta)/(2b)
\ll\min \limits_{g_0}
(|[\mathbf{U}^{0}(t)]_{0g_0}|)$,
%
where
$|[\mathbf{U}^{0}(t)]_{00}|=(1-\sin^2\theta
\sin^2(wt\cos(\theta)/2))^{1/2}$ and
$|[\mathbf{U}^{0}(t)]_{01}|=\sin\theta
|\sin(wt\cos(\theta)/2)|$, with $g_0=0,1$.
Note that the sufficient condition here is stronger than
the necessary one because $5w\sin\theta(|\cos\theta|+\sin\theta)/(2b)$
$\geq$ $w\sin\theta|\cos\theta+\sin\theta|/(2b)$. Moreover,
looking at Eqs.~(\ref{strongerMat})
and (\ref{suf3}), and in particular (\ref{d2}), 
we can show that in general the practical sufficient condition implies the 
necessary one whenever the gap is constant. 
Since for this example the natural choice for the 
perturbative parameter $v$ is the rotating frequency of the field ($v=w$)
\cite{Rig10}, we have $wt\leq 1$ for $t\in[0,T]$. This implies
that $|[\mathbf{U}^{0}(t)]_{00}|\geq |[\mathbf{U}^{0}(t)]_{01}|$ and 
$5w\sin\theta(|\cos\theta|+\sin\theta)/(2b)\geq w^2t\sin^2(\theta)/b$
during the whole evolution of the state. Hence, the sufficient condition
boils down to only one equation, 
$5w\sin\theta(|\cos\theta|+\sin\theta)/(2b)\ll\sin\theta|\sin(wt\cos(\theta)/2)|$,
leading to $5w/(2b)\ll |\sin(wt\cos(\theta)/2)|/(|\cos\theta|+\sin\theta)$.
Note that when $t\approx 0$ and/or $\theta \approx \pi/2$, 
$|[\mathbf{U}^{0}(t)]_{01}|\approx 0$ and we must work with the non-null coefficient
$|[\mathbf{U}^{0}(t)]_{00}|$. In this case the sufficient condition is $5w/(2b)\ll1$.

It is important to remark now that if $w\geq b$ we cannot satisfy the 
sufficient condition, no matter what the value of $\sin\theta$ is. 
Indeed, since both terms appearing at the rhs of the sufficient conditions	
are smaller than one, assuming $w\geq b$ leads to a lhs greater than one. 
The sufficient conditions are then consistent with the cases where
the necessary condition fails. We cannot have $\sin\theta \approx 0$ and
$w\geq b$ as an instance in which DAA holds.

Our last task is to show that for $w<b$ these conditions imply DAA.
In other words, we must use them to show
that the absolute values of the coefficients multiplying
$|1^0(t)\rangle$ and $|1^1(t)\rangle$ of the exact solution are
negligible. Working with the largest of those this is equivalent
to showing that \cite{Rig12} 
$wg(\theta)/b\ll 1$, with
$
g(\theta)=\sin\theta\left(b/\Omega_+
+ b/\Omega_-\right).
$
Using that $g(\theta)$ has a maximum, 
for $\theta \in [0,\pi]$, at $\theta=\pi/2$ given by 
$2b/(b^2+w^2)^{1/2}$
we get
$wg(\theta)/b \leq  2w/(b^2+w^2)^{1/2}
\leq 2w/b.$
Hence, if the sufficient conditions imply that  $2w/b \ll 1$  we are done.
But noting that $|\sin(wt\cos(\theta)/2)|/(|\cos\theta|+\sin\theta)<1/2$ 
the sufficient conditions reduce to $5w/b\ll 1$ which obviously implies
$2w/b\ll 1$. 
 
\textit{Summary.} We established one rigorous necessary condition and two sufficient conditions,
one rigorous and one practical, for the validity of the quantum adiabatic theorem for systems with degenerate 
spectra. Concepts such as ``slowly/adiabatically changing Hamiltonians'' 
and the ``adiabatic approximation'' for degenerate systems, of greatest importance
for the implementation of adiabatic and topological quantum computation as well
as non-Abelian fractional statistics, were quantitatively stated. 
It is this quantitative specification that allows for a precise and rigorous 
formulation of the adiabatic theorem.  Finally, we applied the adiabatic 
theorem 
to an exactly solvable degenerate problem, and provided a complete 
characterization of the mathematical conditions under which the degenerate 
adiabatic approximation
holds.

\begin{acknowledgments}
GR thanks CNPq, FAPESP, and the Brazilian National Institute of
Science and Technology for Quantum Information (INCT-IQ) for funding.
\end{acknowledgments}


\begin{thebibliography}{99}
\bibitem{footnote1} Note that $|\mathbf{\Psi}(t)\rangle^T$
is a ket-row vector while $\langle\mathbf{n}(t)|^T$ is a column-bra
vector \cite{Rig10}.
\bibitem{Rig12} All the details of the calculations here presented as 
well as of those appearing in Ref. \cite{Rig10} will be given elsewhere.
\end{thebibliography}
\end{document}